\documentclass{INTERSPEECH2023}
\usepackage{tabularx,lipsum, multicol, multirow}

\usepackage{cite, amsmath}
\usepackage{xpatch,xcolor}
\usepackage{hyperref}
\usepackage{makecell, booktabs}
\hypersetup{
    colorlinks=true,
    linkcolor=purple,
    filecolor=magenta,      
    urlcolor=purple,
}
\newcommand{\newpara}[1]{\vspace{3pt}\noindent\textbf{#1}}

\interspeechcameraready

\title{Multi-Dataset Co-Training with Sharpness-Aware Optimization\\ for Audio Anti-spoofing}
\name{Hye-jin Shim$^1$, Jee-weon Jung$^{2,\dag}$\thanks{$^\dag$ Work done while author was in Naver Corporation, South Korea.}, Tomi Kinnunen$^1$}
\address{
  $^1$University of Eastern Finland, Finland\\
  $^2$Carnegie Mellon University, USA}
\email{hyejin.shim@uef.fi, jeeweonj@andrew.cmu.edu, tomi.kinnunen@uef.fi}

\begin{document}
\maketitle
\begin{abstract}
Audio anti-spoofing for automatic speaker verification aims to safeguard users' identities from spoofing attacks. Although state-of-the-art spoofing countermeasure(CM) models perform well on specific datasets, they lack generalization when evaluated with different datasets. To address this limitation, previous studies have explored large pre-trained models, which require significant resources and time. We aim to develop a compact but well-generalizing CM model that can compete with large pre-trained models. Our approach involves multi-dataset co-training and sharpness-aware minimization, which has not been investigated in this domain. 
Extensive experiments reveal that proposed method yield competitive results across various datasets while utilizing 4,000 times less parameters than the large pre-trained models.
\end{abstract}

\section{Introduction}
\label{sec:intro}
Automatic speaker verification (ASV) systems~\cite{reynolds2002overview}, even state-of-the-art, have been reported to be easily deceived by \emph{spoofing attacks} including speech synthesis (text-to-speech, TTS), voice conversion (VC).
For the reliable ASV systems, \emph{audio anti-spoofing} has emerged which aims to distinguish the utterance from a real human ({\em bona fide}) or spoofing attacks ({\em spoofed}).
To develop spoofing countermeasure (CM) models, various studies have conducted focusing on feature~\cite{das2019long, yang2019significance, yang2020long}, model architecture~\cite{lai2019assert, li2021replay, tak2021end, jung2022aasist}, and other techniques (e.g. loss function and data augmentation)~\cite{zhang2021one, cohen2022study, tak2022rawboost}.

While most state-of-the-art CMs perform well on specific evaluation datasets, they do not generalize well when cross-evaluated on different datasets~\cite{wang2019cross, das2020assessing, chettri2021data, zhang2021empirical,lu2022acoustic, tak2022automatic, wang2022investigating}.
Several studies have explored improving generalization capability and literature can be divided into two strands:
The first strand exploits techniques to develop a model with feature adjustment, gradient-based methods, and adversarial learning~\cite{das2020assessing, wang2019cross, zhang2021empirical}.
The other strand develops model using domain adaptation, continual learning, and self-supervised learning to leverage large-scale datasets~\cite{ma2021continual, shim2019self, lu2022acoustic, lv2022fake, tak2022automatic, wang2022investigating}.
Even though the latter demonstrated promising results with a large gap, it typically requires additional learning steps with a large, heavyweight model such as wav2Vec2.0~\cite{baevski2020wav2vec} or HuBERT~\cite{hsu2021hubert} that typically contains billions of parameters.

Dating back to the base assumption of large-scale pre-trained models, there is a core premise of ``\textit{more data leads to better (generalization) performance}'' that lies on deep learning as well as statistics.
It expects that exposure of large amounts of data to the model can lead to better generalization. 
Following this idea, abundant studies have demonstrated the effectiveness of training a model on diverse datasets to enhance their robustness and adaptability, namely unsupervised/semi-supervised pre-training and self-supervised learning.
Otherwise, training a model using multiple datasets at once has been well-known as a challenging and unsolved problem because of different characteristics of datasets that may interfere with the target task.
Nevertheless, a few studies have been conducted in this direction~\cite{meletis2018training, kapidis2021multi, zhou2022simple, liang2022multi}.
There is a potential to be improved, so further inspection and exploration yet remain.

\begin{table}[t]
  \caption{Performance comparison of using a single dataset and multiple datasets in audio anti-spoofing. Evaluation is conducted with the ASVspoof 2019 LA dataset.}
  \centering
  \label{tab:motiv}
  \begin{tabularx}{\linewidth}{lc}
    \toprule
    Train Dataset(s) & EER(\%)\\
    \toprule
    ASVspoof 2015 & 38.83 \\
    ASVspoof 2019 & 1.38 \\
    \hline
    ASVspoof 2019 + ASVspoof 2015 & 1.56 \\ 
    ASVspoof 2019 + ASVspoof 2015 + WaveFake & 1.76 \\
    \bottomrule
  \end{tabularx}
  \vspace{-15pt}
\end{table}

In this study, we aim to develop a compact and well-generalized CM model leveraging multi-dataset co-training.
At an early stage of the present study, we conducted pilot experiments with gradually enlarging datasets for training the model at once.
The result is shown in Table~\ref{tab:motiv} and its outcome indicates that merely combining datasets that span different domains does not guarantee generalization.
These observations motivated us to address more elaborate ways to optimize the model when using multiple training datasets.
To mitigate this problem, we explore the way to reduce the perturbation which can be caused by domain mismatch across different datasets.
Recent studies have shown gradient-based methods, especially sharpness-aware related works~\cite{foret2020sharpness, kwon2021asam, liu2022towards}, demonstrate to avoid a severe perturbation during the training process and enhance generalization capability.
Here, the term \textit{sharpness} in this context refers to the curvature of neighborhoods in the loss surface.

To this end, we exploit two recently proposed optimization techniques: \emph{sharpness-aware minimization} (SAM)~\cite{foret2020sharpness} and \emph{Adaptive sharpness-aware minimization} (ASAM)~\cite{kwon2021asam}.
SAM and ASAM are both designed to find flat minimas by taking into account the sharpness in the loss surface in addition to the gradients.
Hence, we hypothesize that combining sharpness-aware training --- an approach designed to avoid sharp loss minima --- with multiple dataset co-training (to handle diverse data) has the potential to lead to improving the generalization of CM.

Our study first attempts to optimize multi-dataset co-training and also practical effectiveness of sharpness-aware training remains presently unknown in the CM task. 
Providing initial answers to this question forms the main novelty of our work;
we implement our proposed method using the state-of-the-art graph neural network-based AASIST~\cite{jung2022aasist} model with various evaluation data.
Our comprehensive experimental validation reveals that both approaches are effective in that the proposed model shows competitive results throughout various datasets using a number of parameters more than 4000 times less than the large pre-trained models and leads to better generalization.


\section{Multi-dataset training}
\label{sec:mdl}

\subsection{Related works}

Several previous works have focused on enlarging the amount of training data for the training based on ``\textit{more data leads to better (generalization) performance}\'' in line with fitting the general distribution with large amounts of data.
Both unsupervised learning and self-supervised learning align with this principle aimed to enhance model generalization with more data.
However, above mentioned studies are taking into account the conditions that are hard to get labeled data for the same task.
It means if labeled data is available, it is basically helpful for training, however, multi-dataset co-training is even unveiled yet compared to the methodologies for exploiting unlabeled data.
Few studies~\cite{zhu2016we, liang2022multi} have conducted multi-dataset co-training in other domains.

There exists a few preliminary works in audio spoofing utilizing multiple datasets ~\cite{paul2017generalization, das2020assessing}. 
They concentrated on developing a single model that can detect diverse types of attacks as audio spoofing attacks can be divided into two categories: logical access (LA) and physical access (PA). 
The former includes TTS and VC attacks, whereas the latter refers to replay attacks only.
However, no research has explored multi-dataset training to deal with one category of attack (either LA or PA).
The potential effectiveness of training the model with multiple datasets simultaneously for the same task has yet to be explored in depth, but it would be worthwhile to investigate further.

\subsection{Summary of datasets used in this study}
We use three datasets concurrently to train a \emph{single model} in a \emph{single phase}: 
ASVspoof 2015~\cite{wu2014asvspoof},  ASVspoof 2019 LA~\cite{todisco2019asvspoof}, and WaveFake ~\cite{frank2021wavefake}.
The latest ASVspoof edition in 2021 additionally introduced DeepFake (DF) scenario which includes lossy codecs used for media storage.
In this study, we \emph{only} deal with LA spoofing attacks for training a model.
To address generalization to an unseen domain, we use the ASVspoof 2021 LA and DF tasks for the evaluation.
Note that ASVspoof 2015, ASVspoof 2019, and a part of ASVspoof 2021 are based upon the Voice Cloning Toolkit (VCTK) corpus~\cite{vctk}; however, they cover different attacks.
Supporting this basis, it is assumed that ASVspoof 2015 evaluation can be theoretically easy when a CM is trained on the more diverse LA 2019 train set.
However, it has empirically confirmed that this is not the case~\cite{das2020assessing, wang2022investigating}.
An overview of the selected datasets is shown in Table ~\ref{tab:data_summary}. 

\newpara{ASVspoof 2015}~\cite{wu2014asvspoof} is the earliest and smallest database among the four existing ASVspoof editions.
The evaluation set consists of five known and five unknown attacks composed of different TTS and VC systems. 
In this context, the term \textit{known} attack indicates an attack in the train and test set is overlapped, while \textit{unknown} attack indicates to scenarios where the test set includes attacks that were not encountered during the training phase.

\newpara{ASVspoof 2019}~\cite{todisco2019asvspoof} is a large-scale dataset that covers advanced technologies developed during the four years following ASVspoof 2015.
It includes 6 and 13 types of spoofing attacks in train and test, respectively.
There are two known attacks, four partially known attacks, and seven unknown attacks between train and test sets.
Here, \textit{partially known} attack denotes a scenario where some of the attacks are present in both the train set and test set, but some attacks are not present in the train set.

\newpara{WaveFake}~\cite{frank2021wavefake} is collected using six different state-of-the-art TTS methods. 
It considers the created samples to resemble the training distributions. 
All spoofed data has been generated using the last, competitive VC and TTS models.
Note that we utilize whole spoofed speech for WaveFake since no standardized test protocol exists.
WaveFake contains spoofed utterances from two speakers, originating from LJSPEECH~\cite{ljspeech17} and JSUT~\cite{DBLP:journals/corr/abs-1711-00354} datasets, respectively.

\newpara{ASVspoof 2021}~\cite{ASV2021challenge} is the latest and hardest edition of the ASVspoof challenge series.
It only contains a test set and introduces real telephony systems both applied to encoding and transmission artifacts in the LA scenario of the ASVspoof 2019 dataset.
The DF scenario additionally consists of bona fide and spoofed data processed through various audio compressors containing data from two additional datasets, VCC 2018~\cite{lorenzo2018voice} and VCC 2020~\cite{zhao2020voice}.

\begin{table}[t]
  \caption{Summary of data statistics used in this study. \#Spks, \# Utts, and \# Conds refer to the number of speakers, utterances, and spoofing conditions, respectively. Division of train and test set is indicated by /. }
  \centering
  \label{tab:data_summary}
  \begin{tabularx}{\linewidth}{lccc}
    \toprule
    Dataset & \# Spks & \# Utts & \# Conds\\
    \toprule

    ASVspoof 2015 & 25 / 46 & 16375 / 193404 & 5 / 10\\
    ASVspoof 2019 LA & 20 / 48 & 25380 / 108978 & 6 / 13\\  
    WaveFake & 2 &  117985 & 6\\
    \hline
    ASVspoof 2021 LA & - / 48 & 181566 & - / 13 \\
    ASVspoof 2021 DF & - / 48 & 611829 & - / 13 \\
    \bottomrule
    \end{tabularx}
    \vspace{-15pt}
\end{table}

\section{Sharpness-aware optimizations}
\label{sec:sam}
When working with multiple datasets simultaneously, a model may easily struggle with domain information between the main task, despite having access to explicit class labels.
Domain information can distract the model from converging, although it may generalize better once the train loss has well converged.
We thus seek methods that can prevent the model from being distracted by domain discrepancies between different datasets.
Besides, this direction can also cast away the necessity of the additional pre-training and fine-tuning steps in a straight way.

\begin{table*}[]
  \caption{In-domain and out-domain evaluation results for ASVspoof 2015, ASVspoof 2019 LA, ASVspoof 2021 LA \& DF. For model training, ASVspoof 2015, ASVspoof 2019 LA, and WaveFake are used. The first row and column refer to training and evaluation datasets, respectively. The boldface in the last row and the last column indicate the best results between training objectives and training datasets, respectively. On the other hand, the underline results refer to the best results in each evaluation set.}
  \centering
  \label{tab:pooled}
  \setlength\tabcolsep{5pt}
\begin{tabular}{lccccccccccccc}
\hline
                                                                        & \multicolumn{3}{c}{\textbf{ASVspoof 2015}}                                                                         & \multicolumn{3}{c}{\textbf{ASVspoof 2019 LA}}                                                                      & \multicolumn{3}{c}{\textbf{ASVspoof 2021 LA}}                                                                      & \multicolumn{3}{c}{\textbf{ASVspoof 2021 DF}}                                                                      & \multicolumn{1}{l}{}               \\ \cline{2-13}
\multicolumn{1}{c}{\textbf{}}                          & \multirow{2}{*}{\begin{tabular}[c]{@{}c@{}}w/o\\ SAM\end{tabular}} & \multirow{2}{*}{SAM} & \multirow{2}{*}{ASAM} & \multirow{2}{*}{\begin{tabular}[c]{@{}c@{}}w/o\\ SAM\end{tabular}} & \multirow{2}{*}{SAM} & \multirow{2}{*}{ASAM} & \multirow{2}{*}{\begin{tabular}[c]{@{}c@{}}w/o\\ SAM\end{tabular}} & \multirow{2}{*}{SAM} & \multirow{2}{*}{ASAM} & \multirow{2}{*}{\begin{tabular}[c]{@{}c@{}}w/o\\ SAM\end{tabular}} & \multirow{2}{*}{SAM} & \multirow{2}{*}{ASAM} & \multirow{2}{*}{\textit{Average}} \\
\multicolumn{1}{c}{}                                                        &                                                                    &                      &                       &                                                                    &                      &                       &                                                                    &                      &                       &                                                                    &                      &                       &                                    \\ \hline
\textbf{2015}                                                               & 8.25                                                               & 6.50                 & 5.83                  & 38.83                                                              & 29.50                & 30.70                 & 39.87                                                              & 32.27                & 31.09                 & 33.54                                                              & 28.40                & 21.80                 & 25.55                              \\ \hline
\textbf{2019 LA}                                                            & 5.98                                                               & 4.32                 & 3.53                  & 1.38                                                               & 1.06                 & 1.48                  & 12.18                                                              & \underline{7.08}                 & 10.18                 & \underline{18.20}                                                              & 21.16                & 19.58                 & 8.84                               \\ \hline
\textbf{\begin{tabular}[c]{@{}l@{}}2015 \\ + 2019 LA\end{tabular}}           & 0.69                                                               & 0.80                 & 0.83                  & 1.56                                                               & 1.07                 & \underline{0.99}                  & 12.25                                                              & 12.52                & 9.37                  & 21.02                                                              & 26.32                & 22.14                 & 9.13                               \\ \hline
\textbf{\begin{tabular}[c]{@{}l@{}}2015\\ +2019 LA\\ +WaveFake\end{tabular}} & 0.82                                                               & \underline{0.66}                 & 1.82                  & 1.76                                                               & 1.78                 & 1.27                  & 11.05                                                              & 13.08                & 12.41                 & 19.76                                                              & 19.49                & 19.84                 & \textbf{8.65}                      \\ \hline
\multicolumn{1}{c}{\textit{Average}}                                       & 3.94                                                               & 3.07                 & \textbf{3.00}         & 10.88                                                              & \textbf{8.35}        & 8.61                  & 18.84                                                              & 16.24                & \textbf{15.76}        & 23.13                                                              & 23.84                & \textbf{20.84}        &                                    \\ \hline
\end{tabular}
\vspace{-10pt}
\end{table*}


In particular, \emph{sharpness-aware minimization} (SAM)~\cite{foret2020sharpness} recently demonstrates state-of-the-art performance in various tasks~\cite{bahri2021sharpness, andriushchenko2022towards, coldenhoff2022model, caldarola2022improving}.
SAM aims to find a \emph{flat} region in the parameter space with both low loss itself and neighborhoods, seeking flat minima.
It uses worst-case perturbation of the model parameters on every iteration in the training phase and can be easily implemented on top of existing optimizers such as Adam\cite{kingma2014adam}.
Moreover, there are several follow-up studies related to sharpness.
For instance, the authors in ~\cite{kwon2021asam} proposed a scale-invariant version, \emph{adaptive} variant of SAM, adaptive SAM (ASAM).
It solves the scale dependency problem by removing the effect of scaling and helps build a solid correlation with the generalization gap.
In the following subsections, we detail both SAM and ASAM.



\subsection{Sharpness-Aware Minimization}

Given a labeled training set $S = \{(x_1, y_1), \dots, (x_n, y_n)\}$ drawn from i.i.d an unknown data distribution $D$, the \emph{training loss} function with model parameter $\textbf{w}$ and \emph{population loss} are defined as $L_S$ and $L_D(w)\fallingdotseq \mathbb{E}_{(x,y)~D}[l(\textbf{w},x,y)]$, respectively.
While the training loss is the empirical/sample-based estimator, the population loss refers to the corresponding theoretical quantity when the knowledge of the actual joint distribution $(x, y)$ is fully known.
So, population loss can be thought of as the empirical training loss, applied to an infinitely large training set $(n \rightarrow \infty)$.

Our goal is to select $\textbf{w}$ not only for having low training loss $L_S(\textbf{w})$ but also for low population loss $L_D(s)$, for improved generalization. 
To achieve such a goal, SAM is designed to minimize the following PAC-Bayesian generalization upper bound, where the sharpness term appears explicitly in another following equation. (The term in brackets indicates the sharpness of $L_s$ at $\textbf{w}$):

\begin{gather*}
L_D (\textbf{w}) \leq \max_{\lVert \varepsilon \lVert_2 \leq \rho} L_S(\textbf{w}+\varepsilon)+h(\lVert \textbf{w} \lVert^2_2 / \rho^2) \\
=[\max_{\lVert \varepsilon \lVert_2 \leq \rho} L_S(\textbf{w}+\varepsilon)-L_S(\textbf{w})] + L_S(\textbf{w})+h(\lVert \textbf{w} \lVert^2_2 / \rho^2)
\end{gather*}

Here, $h$ is a strictly increasing function under conditions on $L_D(\textbf{w})$and $\rho$ is a predefined constant controlling the radius of a neighborhood in $l^p$ ball ($p \ni \left[ 1, \infty \right]$, \cite{foret2020sharpness} revealed that $p=2$ is optimal).
A detailed explanation of the PAC-Bayesian generalization bound is omitted for the limited space. 
Refer to Appendix A.1 of ~\cite{foret2020sharpness}) for full details. 

Finally, for any $\rho>0$ and $\varepsilon\approx0$ (to avoid division by 0), the model loss is defined as:

\begin{gather*}
\min_\textbf{w} L^{\text{SAM}}_S (\textbf{w}) + \lambda\lVert \textbf{w} \lVert^2_2, \\
\text{where} L^{\text{SAM}}_S (\textbf{w}) \triangleq \max_{\lVert \varepsilon \lVert_p \leq \rho} L_S(\textbf{w}+\varepsilon)
\end{gather*}

\begin{table}[]
  \caption{Per-attack results of ASVspoof 2021 DF evaluation. The best results are selected between w/o SAM, SAM, and ASAM. The best results in the same row are represented in boldface. ((a) ASVspoof 2015, (b) ASVspoof 2019, (c) WaveFake.)}
  \centering
  \label{tab:per-attack}
  \begin{tabularx}{\linewidth}{lcccc}
    \toprule
    Attack &
    (a) & (b) & (a)+(b) & (a)+(b)+(c)\\
    \toprule
    Traditional & 21.54 & 12.18 & 14.18 & \textbf{10.77}\\
    Wav.Concat. & 55.22 & \textbf{12.07} & 20.32 & 13.09  \\
    Neural AR & 46.82 & \textbf{23.10} & 27.70 & 24.87\\
    Neural non-AR & 40.23& \textbf{20.47} & 25.05 & 23.21\\
    Unknown & 24.45 & 20.38 & 17.94 & \textbf{16.05}\\
    \hline
    Pooled & 33.54 & 18.20 & 22.14 & 19.49\\
    \bottomrule
  \end{tabularx}
  \vspace{-15pt}
\end{table}

\subsection{Adaptive Sharpness-Aware Minimization}
Even if the vanilla SAM performs usually well, it is easily affected by parameter re-scaling as a sharpness term in SAM defined on a rigid region with a fixed radius.
This may disturb the generalization of SAM.

To address this shortcoming, ASAM~\cite{kwon2021asam} uses adaptive sharpness that removes the effect of scaling and adjusting the maximization region, leading to an improved training path.
It utilizes the normalization operator $T^{-1}_\mathbf{w}$ and achieves better generalization compared to SAM.
Firstly, the normalization operator of weight $\textbf{w}$ can be defined, if $T_\textbf{w}$ is a family of invertible linear operators and $T^{-1}_A\textbf{w}$ $=T^{-1}_\textbf{w}$ for any invertible scaling operator $A$, which does not alter  the loss function.
With this normalization operator, adaptive sharpness objective function is defined as:

\begin{align*}
L^{\text{ASAM}}_S (\textbf{w}) \triangleq \max_{\lVert T^{-1}_\mathbf{w} \epsilon \lVert_p \leq \rho} L_S(\textbf{w}+\varepsilon)
\end{align*}

\section{Experimental settings}
For experiments, we deploy the ``light'' version of the recent AASIST model~\cite{jung2022aasist}, referred to as AASIST-L.
It includes a graph attention layer to capture information both in spectral and temporal domains and max graph operations to select features in a competitive manner.
The main difference between AASIST and AASIST-L is the number of parameters for practical purposes including 297K and 85K numbers of parameters, respectively.
We use Adam~\cite{kingma2014adam} as our base optimizer.
When exploiting SAM and ASAM, optimization proceeds similarly, but with the additional sharpness term added to the training loss as explained above.
As our aim is to focus on the generalization throughout corpora rather than dealing with architectural details, we did not adjust parameters (e.g. learning rate, pooling ratio).
Full details of AASIST model can be referred to ~\cite{jung2022aasist}.
All models were implemented using PyTorch and trained for 100 epochs.
Performance evaluation is done based on equal error rate (EER) and we selected the best-performing model in terms of EER on the development set. 
Code for experiments of this study is available in: \url{https://github.com/shimhz/MDL_sharpness}. 

\begin{table}[]
  \caption{The comparison of mini-batch composition strategies using ASVspoof 2019 LA evaluation. Pooled mini-batch refers to the condition which ignores the balance between datasets, and balanced mini-batch refers to the condition which considers the balance between multiple datasets.}
  \centering
  \label{tab:batch}
  \begin{tabularx}{\linewidth}{lccr}
    \toprule
    Training datasets & Mini-batch & Loss & EER \\ \hline
2015 + 2019 LA & pooled & w/o SAM & 1.56 \\ \hline
2015 + 2019 LA & \textbf{balanced} & w/o SAM & 1.49 \\ \hline
\multirow{2}{*}{\begin{tabular}[c]{@{}c@{}}2015 + 2019 LA \\ + WaveFake\end{tabular}} & \multirow{2}{*}{pooled} & \multirow{2}{*}{ASAM} & \multirow{2}{*}{1.27} \\
 &  &  &  \\ \hline
\multirow{2}{*}{\begin{tabular}[c]{@{}c@{}}2015 + 2019 LA \\ + WaveFake\end{tabular}} & \multirow{2}{*}{\textbf{balanced}} & \multirow{2}{*}{ASAM} & \multirow{2}{*}{1.09} \\
 &  &  &  \\
    \bottomrule
  \end{tabularx}
  \vspace{-15pt}
\end{table}

\section{Results and Analyses}
\label{ssec:sam}

\newpara{Main results}
In Table~\ref{tab:pooled}, we validate our proposed model using both in-domain datasets (ASVspoof 2015 and ASVspoof 2019 LA) and out-of-domain datasets (ASVspoof 2021 LA and ASVspoof 2021 DF).
Firstly, as for the multi-dataset co-training, the pooled results are shown in the last column. 
We confirm that the best result achieved when all three datasets are all utilized.
Even though the full usage of three datasets did not show a consistent result as the best result, the effectiveness of using multiple datasets was demonstrated.

Secondly, the results of sharpness-aware optimizations can be easily comparable in the last row.
Sharpness-aware optimization methods improve the performance in most cases, regardless of whether using a single dataset or multiple datasets.
Except for the ASVspoof 2019 LA evaluation case that gets the lowest EER using SAM, ASAM shows the best results.
Through results, we observe that SAM and ASAM substantially benefit the model optimization.

\newpara{Per-attack results on ASVspoof 2021 DF} 
For further analysis, Table ~\ref{tab:per-attack} shows the results for each attack in ASVspoof 2021 DF evaluation result.
We depict the best results for each training dataset combination
: each column refers to ASVspoof 2015 (w/ ASAM), ASVspoof 2019 (w/o SAM), ASVspoof 2015 + ASVspoof 2019 (w/o SAM), and ASVspoof2015 + ASVspoof 2019 + WaveFake (w/ SAM). 
The interesting thing to be noted is that (a)+(b)+(c) which includes all datasets show superior in Traditional and Unknown attack in a large gap compared to the best result which is from (b) only.  
In terms of generalization, \emph{unknown} is the most critical subset; thus we interpret that the lowest EER in unknown signifies better generalization. 
Hence, these results back up the effectiveness of the proposed method for the generalization performance.

\newpara{Mini-batch composition} When utilizing multiple datasets, an imbalance exists between different datasets. We thus further explore whether balancing the number of samples drawn from each dataset can be advantageous in terms of performance. Table~\ref{tab:batch} describes the results, which confirm that simply balancing the samples between different datasets within a mini-batch are helpful. We confirm improvement in both reported train dataset configurations, where the performance of the most extensive setting is further improved by 14\% relative.

\newpara{Comparison with other studies} In Table ~\ref{tab:comparison}, we compare our results with other state-of-the-art systems including the studies utilizing a large pre-trained model.
Among four evaluation protocols, our model demonstrates competitive performance with only 85K parameters in two protocols: ASVspoof 2015 and ASVspoof 2019 LA.
In the other two remaining protocols, our model underperforms; nonetheless, taking into account the number of parameters and training time, we argue that our approach remains competitive.
Given the fact that the purpose of CM models is to aid ASV systems, lightweight yet well-generalizing models are worth further investigation.

\begin{table}[t]
  \caption{Comparison with other state-of-the-art results including the research which utilized pre-trained model with large datasets.}
  \centering
  \label{tab:comparison}
  \begin{tabularx}{\linewidth}{lcc}
    \toprule
    \textbf{Method} & \textbf{\# Params} & \textbf{EER(\%)}\\
    \toprule
    \multicolumn{3}{c}{\textbf{ASVspoof 2015}}\\
    \toprule
    Primary result~\cite{wu2015asvspoof} & - & 1.21 \\
    \textbf{Ours} & 85K  & 0.66\\
    \hline
    \multicolumn{3}{c}{\textbf{ASVspoof 2019 LA}}\\
    \toprule
    SSAD+LCNN big~\cite{jiang2020self} & - & 5.31\\
    Imag-pre + J\&S~\cite{lu2022acoustic} & - & 0.87\\
    wav2Vec2.0~\cite{wang2022investigating}  & 317M + (290±30K) & 1.28 \\
    HuBERT-XL~\cite{wang2022investigating}  & 317M +  (290±30K) & 3.55 \\
    \textbf{Ours} &  85K  & 0.99\\
    \hline
    \multicolumn{3}{c}{\textbf{ASVspoof 2021 LA}}\\
    \hline
    Img-pre+RawBoost~\cite{lu2022acoustic} & - & 7.71\\
    wav2Vec2.0-XLSR~\cite{tak2022automatic} & 317M + 297K & 6.15\\
    wav2Vec2.0-XLSR~\cite{wang2022investigating}  & 317M + 297K & 9.66\\
    HuBERT-XL~\cite{wang2022investigating}  & 317M +  (290±30K) & 9.55 \\
    \textbf{Ours} & 85K  & 7.08\\
    \hline
    \multicolumn{3}{c}{\textbf{ASVspoof 2021 DF}}\\
    \hline
    Img-pre+RawBoost~\cite{lu2022acoustic} & - & 19.11\\
    wav2Vec2.0-XLSR~\cite{tak2022automatic}  & 317M + 297K & 7.69\\
    wav2Vec2.0-XLSR~\cite{wang2022investigating}  & 317M + 297K & 4.75\\
    HuBERT-XL~\cite{wang2022investigating}  & 317M +  (290±30K) & 13.07 \\
    \textbf{Ours} & 85K  & 18.20\\
    \bottomrule
    \end{tabularx}
    \vspace{-15pt}
\end{table}

\section{Conclusions}
Recent studies have widely exploited large pre-trained models to leverage as much data as possible to develop a well-generalized model.
While training a single model with multiple datasets is a straightforward way to utilize diverse data without additional training, it is well-known that handling domain differences is challenging. Given the nature that CM models are inherently built to support ASV systems, these enormous systems can be potentially not applicable because of their size. In this paper, we explore a case study in the audio anti-spoofing field which lacks a large amount of data compared to other research domains. To optimize the model to handle multiple datasets simultaneously, we utilize sharpness-aware methodologies, which include a curvature-based term in the objective function to reduce the gap between the variance of the data. Using a number of parameters more than 4000 times less than the large pre-trained models, our proposed method demonstrates effectiveness in both in-domain evaluations on unknown attacks and out-of-domain evaluations. 

\section{Acknowledgements}
This work was supported by the Academy of Finland (Decision No. 349605, project ``SPEECHFAKES'')

\bibliographystyle{IEEEtran}
\bibliography{shortstrings,mybib}

\end{document}